\documentstyle[12pt]{article}

\newcommand{\Section}[1]{\section{#1}}

       \def\C{\rm {I\kern-.520em C}}

           \def\be{\begin{equation}}
           \def\bea{\begin{eqarray}}
           \def\ba{\begin{array}}
               \def\ee{\end{equation}}
               \def\eea{\end{eqnarray}}
               \def\ea{\end{array}}
                   \def\os{osp(2\mid 1)}
                   \def\st{sl(2)}
                   
              \def\uos{U_q(osp(2\mid 1))}

                       \def\d{\partial}
                       \def\D{D}
                       \def\M#1#2{M^{#1}_{#2}}
                       \def\T#1#2{T^{#1}_{#2}}
                       \def\t{\theta}
                       \def\dt#1{\d_{\t_{#1}}}
                       
                       \def\zz{<\Phi(z_1,\t_1)\Phi(z_2,\t_2)>}

                       \def\q#1{q^{{#1}h}}

                           \def\s{\sum\limits}
                           
                           \def\p{\prod}
                      \def\f{\frac}
                     \def\tp{\otimes}
                               \def\qq{\frac{1}{q-q^{-1}}}

                                   \def\rar{\rightarrow}

\begin{document}

\begin{titlepage}
\begin{flushright}
                                 hep-th/9503191\\
                                 IPM-95-078\\
                                 SUTP/1/95/73 \\
                                             \end{flushright}
\vspace{-12pt}
\begin{center}
\begin{large}
             {\bf FREE FIELD REPRESENTATION OF $\os$ AND
             $\uos$ AND N=1 (q-)SUPERSTRING CORRELATION FUNCTIONS}
\end{large}

\

                          {\bf W.-S. Chung${}^*$ and  A.
Shafiekhani${}^\dagger$}
\footnote{e-mail address:
ashafie@physics.ipm.ac.ir}\\
\vspace{12pt} {\it
                ${^*}$Theory Group, Dept. of Physics\\
                College of Natural Sciences,\\
                Gyeongsang National University, Jinju, 660.701, Korea\\
\vspace{8pt}
${^\dagger}$Institute for Studies in Theoretical Physics and Mathematics\\
             P.O.Box: 19395-5746, Tehran, Iran,\\
          and${\;\;}$   Dept. of Physics,
                 Sharif University of Technology,\\
                 P.O. Box: 11365-9161, Tehran, Iran
}\\
\end{center}
\abstract{

The free field realization of irreducible representations of $\os$ is
constructed, by a unified and systematic scheme. The $q$-analog of
this unified scheme is used to construct $q$-free field realization
of irreducible representations of $\uos$. By using these realization, the
two point function of $N=1$ superconformal (q-superconformal) model based on
$\os$ ($\uos$) symmetry have been calculated.

\vfill
}

\end{titlepage}
{\Section {Introduction}}

In modern theoretical physics, free field realizations of Lie
algebras drive an important role from technical point of view. Usually,
this simplification allows to go much further with any kind of formal
manipulations.

Free field representations of Lie (super)algebras and quantum
(super)algebras [1-8] are of interest because, like quantum
algebras they have many applications in (super)conformal field
theories \cite{cft}, inverse problems and integrable systems \cite{andre,frt}
and statistical mechanics \cite{sm}.

$\os$ as a Lie superalgebra plays a special rule, in part analogous to that
played by $\st$ in simple Lie algebras. It is therefore of particular
interest to construct the free field representation of this superalgebra.
A natural question is whether such a realization exists in the case of
$\uos$ too.

There have been attempts to construct for a Heisenberg representation of
$\os$ and $\uos$
in ref. \cite{bgt}, by Schwinger method. Due to the method of construction,
they are not generalizable to all irreducible representations and
affinization.

There has been a lot of interest on class of superconformal models specially
super-WZW models based on nonablian nonsemisimple
Lie groups [14-19], particularly because they allow the construction
of exact string background. One class of these superconformal models is
N=1 superconformal with
$\os$ super-Virasoro symmetry. Correlation functions for such theories had
been calculated by different methods \cite{q,ks}. In this letter
we give a very simple way of calculation, which is extendable to all other symmetrice
groups. The same question arise if there
exists any q-superconformal symmetry for such a theory, what will be then the
q-analog correlation functions?

Our aim in this letter is to apply the unified and systematic scheme given in
ref. \cite{aziz} for $A_n$ series, for $\os$ and $\uos$ and use these realizations
to calculate two point function of $N=1$ super(q-super)conformal model based on
$osp(2|1)(U_q(osp(2|1))$ symmetry group.

The structure of this letter is as follows: In section 2, we construct
the differential realization of $\os$ for all irreducible
representations.
In section 3, we use the same method of section 2 to present $q$-difference
operator realization of $\uos$, for any irreducible
representation. In section 4, we will use the results of previous sections to
calculate two point function for $N=1$ super(q-super)conformal model based on
$\os$ and $\uos$ symmetry group.
\noindent
{\Section {Free Field Representation of  $\os$}}

$\os$ is a rank one Lie superalgebra. This algebra has three even
generators $X_\pm$, $H$ of $sl(2)$ and two odd generators of $V_\pm $, which
satisfy the following (anti)commutation relations:
\be \label{osc}
\begin{array}{c}
      \{ V_+, V_- \} =H, \hspace{2cm} \{ V_\pm , V_\pm \}=2X_\pm \\

 [ X_+ , X_- ]= H, \hspace{1cm} [H, X_\pm ]=\mp 2X_\pm , \hspace{1cm}
 [H, V_\pm ]=\mp V_\pm.
\end{array}
\ee
Let $2h$ be any arbitrary highest weight,
\be
       V_+{\mid}{2h}>=0, \hspace{2cm} H{\mid}{2h}>=2h{\mid}2h>.
\ee
$\{V_+\}$ and $\{H\}$ are the isotropic subalgebras of state $\mid2h>$.
According to ref. \cite{aziz} the states in representation space are
\be\label{state}
    e^{\t V_-} e^{zX_-}{\mid}{2h}>
\ee
where $\t$ and $z$ are Grassmanian and complex variables, respectively.
The basis vectors  of this vector space are
$\{{\mid}{2h}>,V_-{\mid}{2h}>, ..., (V_-)^{4h}{\mid}{2h}>\}$.

By the action of $V_\pm$ and $H$ on (\ref{state}) and using the above
(anti)commutation relations we will have:
$$
V_-e^{\t V_-}e^{zX_-}{\mid}{2h}>=V_-(1+\t V_-)e^{zX_-}{\mid}{2h}>=
(\d_\t+\t\d_z)e^{\t V_-} e^{zX_-}{\mid}{2h}>
$$
\be
      V_+e^{\t V_-}e^{zX_-}{\mid}{2h}>=(2h\t+z\d_\t+z\t\d_z)
      e^{\t V_-}e^{zX_-}{\mid}{2h}>
\ee
$$
    He^{\t V_-}e^{zX_-}{\mid}{2h}> = (2h+\t\d_\t +2z\d_z)
    e^{\t V_-}e^{zX_-}{\mid}{2h}>.\nonumber \\
$$
Let us define
\be\label{os1}
\ba{c}
    v_+=\d_\t +\t\d_z \\
v_-=\t2h+z\d_\t+z\t\d_z \\
H=2h +\t\d_\t+2z\d_z
\ea
\ee
where $\d_z=\frac{\d}{\d z}$, $\d_\t=\frac{\d}{\d_\t}$,
$[\d_z,z]=1 $, and $\{\d_\t,\t\}=1$. One can consider
this, as the representation
of $\os$ on the super-sub-space of analytic functions spanned by the monomials,
$\{1,z,z^2,...,z^{4h}, \t, \t z,\cdots, \t z^{4h}\}$.
We will find that
this new operator realization,
which satisfies the algebra of (\ref{osc}),
is a finite-dimensional irreducible representation of $\os$.

Up to now, everywhere the covariant derivative in superspace was being defined
by
\be
D:=\d_\t+\t\d_z.
\ee
As we can see this is not definition but it is differential
realization of $v_+$ on superspace of analytical function of $f(z,\t)$.

Geometrically, this realization describes the right action of the group on
sections of a holomorphic line bundle over the flag manifold $\os /T$,
where $T$ is isotropic subalgebras of state $\mid 2h >$.

Similar results for the left action on flag manifold with states on
representation space,
$$<-2h\mid e^{z X_-}e^{\t V_-},\hspace{0.5cm} <-2h\mid X_+=0, \hspace{0.5cm}
<-2h\mid H=<-2h\mid(-2h),
$$
will obtain as follows
\be\label{os2}
\ba{c}
v_-=\d_\t +\t\d_z \\
v_+=-z\d_\t -\t z\d_z -2h\t \\
H=-2z\d_z-\t\d_\t -2h.
\ea
\ee
\noindent
{\Section { $\uos$}}

Let us first fix the notations. Consider $q$-exponential function:
\be
e_q^x:=\s^\infty_{n=0}\f{x^n}{[n]!},\hspace{0.25cm}
[n]:=\f{q^{n}-q^{-n}}{q-q^{-1}}
\ee
and the $q$-difference operator
\be
         D_xf(x)=\f{f(qx)-f(q^{-1}x)}{(q-q^{-1})x}=
         \f{1}{(q-q^{-1})x}(M^{+1}_x -M^{-1}_x)f(x)
\ee
where $M_x$ is a translation operator, defined by $M_x^nf(x)=f(q^nx)$.

Just as in the case of ordinary $\os$ for right action on flag manifold,
we take the states,
$e_q^{\t V_-}e^{zX_-}_q \mid 2h>$,
in the representation space. The algebra relations are \cite{ch}:
\be \label{uosc}
\begin{array}{c}
      \{ V_+, V_- \} =[H]_q, \hspace{0.5cm} \{ V_\pm , V_\pm \}=2X_\pm
      ;\hspace{0.5cm}[H]_q:=\f{q^H-q^{-H}}{q-q^{-1}}\cr

 [ X_+ , X_- ]= H, \hspace{1cm} [H, X_\pm ]=\mp 2X_\pm , \hspace{1cm}
 [H, V_\pm ]=\mp V_\pm
\end{array}
\ee
with the same procedure as ordinary case one will find the following
$q$-difference operator realization for any arbitrary highest weight
representation:
\be
\ba{c}
v_+=\D_\t+\t\D_z \\
v_-=\f{q}{q+1}z(q^{2h}M^{+1}+q^{-2h-1}M^{-1})D_\t
+\f{1}{q-q^{-1}}(q^{2h}M^{+2}-q^{-2h}M^{-2})\t\cr
+\f{q}{(q+1)(q-q^{-1})}(q^{2h-1}-q^{-2h}+q^{-2h}M^{-2}-q^{2h-1}M^{+2})\t\cr
H=2h+\t\d_\t+2z\d_z
\ea
\ee
where
\be
\begin{array}{c}
         M^{\pm n}_\t:=T^{\pm n}=q^{\pm n \t\d_\t}
         \sim 1+ (q^{\pm n}-1)\t\d_\t, \hspace{0.5cm} M^{\pm n}_z:=M^{\pm n}
         =q^{z\d_z}\cr
[D_z,z]=\f{q\M{+1}{}+\M{-1}{}}{q+1},\hspace{0.5cm}
\{ D_\t,\t \}=\f{\T{+1}{}+ q\T{-1}{}}{q+1}.
\end{array}
\ee

Similar results for the left action on flag manifold with states on
representation space, $<-2h\mid e_q^{z X_-}e_q^{\t V_-}$, is
\be\label{uos2}
\begin{array}{c}
v_-=\D_\t+\t\D_z \cr
v_+=-\f{q}{q+1}z(\q{2}\M{+1}{}+q^{-2h-1}\M{-1}{})D_\t
-\qq(\q{2}\M{+2}{}-\q{-2}\M{-2}{})\t\cr
-\f{q}{(q+1)(q-q^{-1})}(q^{2h-1}-q^{-2h}+q^{-2h}M^{-2}-q^{2h-1}M^{+2})\t\cr
H=-\t\d_\t-2z\d_z-2h.
\end{array}
\ee
We find that
these new operator realization which satisfy the algebra of (\ref{uosc}),
is a finite-dimensional irreducible representation of $\uos$.

In the limit of $q\to 1$ all the above realization goes to that of the ordinary
$\os$.
\noindent
{\Section {Corrolation functions of $N=1$ super(q-super)conformal models
based on $\os$ and $\uos$}}

The String theory which is invariant under the global superconformal group,
$OSP(2|1)$, generated by $G_{\pm 1/2}$, $L_{\pm 1}$ and $L_0$, is $N=1$
superstring theory.
In what follows we set
$$
v_{\pm}=G_{\pm 1/2},\hspace{0.5cm} H=2L_0
$$
just to use the common notations for Virasoro algebra.

As we know the generators of super-Virasoro algebra of $\os$, $G_{\pm 1/2}$, $L_{\pm 1}$
and $L_{0}$, satisfy the following algebraic relations:
\be\label{vir}
\begin{array}{c}
[L_{+1},L_{-1}]=2L_{0}\hspace{0.5cm}
[L_{0},L_{\pm 1}]=\mp L_{\pm}\cr
\{G_{\pm 1/2},G_{\pm 1/2}\}=2L_{\pm}\hspace{0.5cm}
\{G_{1/2},G_{-1/2}\}=2L_{0}\cr
[L_{0},G_{\pm 1/2}]=\mp G_{\pm 1/2}\\
\end{array}
\ee
In the above algebra, $L_{\pm 1}$ are bosonic generators and $G_{\pm 1/2}$
are fermionic generators.

In order to construct the Ward identity, we take following "co-product"
and differential realization of (\ref{os2}) with $H=2L_0$, to eliminate
the two point function,
which coincide with \cite{bpz}:
\be
\Delta(g)=g\tp I+I\tp g;\hspace{0.5cm} g\in osp(2|1)
\ee
where $I$ means the identity operator. Such a written form will be more visible
in the case of $q$-analog.

Consider the quasi-primary superfield,
$\Phi(z,\t)=\phi(z)+\t f(z)$,
where $\phi(z)$ and $f(z)$ are bozonic and fermionic fields respectively.

Then the Ward identity is given by
\be
\begin{array}{c}
\Delta L_0\zz=(\s^{2}_{i=1}(-z_i\d_{z_i}-\f{1}{2}\t_i\d_{\t_i}-h_i
)\zz=0\cr
\Delta G_{-1/2}\zz=
(\s^{2}_{i=1}(\d_{\t_i} +\t_i\d_{z_i)})\zz=0\cr
\Delta G_{+1/2}\zz=
(\s^2_{i=1}(z_i\d_{\t_i} +\t_i z_i\d_{z_i} +2h_i\t_i))
\zz=0\cr
\Delta L_{-}\zz=(\s^2_{i=1}\d_{z_i})\zz=0\cr
\Delta L_{+}\zz=(\s^2_{i=1}(z_i^2\d_{z_i}+\t_iz_i\d_{\t_i}+h_iz_i))
\zz=0
\end{array}
\ee
where $h_1$ and $h_2$ are conformal weight of $\Phi(z_1,\t_1)$ and
$\Phi(z_2,\t_2)$ respectively.
If we solve the above Ward identity, we have the following expression for two
point function which is well-known \cite{q}.
\be
\zz\sim  (z_1-z_2-\t_1\t_2)^{-(h_1+h_2)}=z_{12}^{-(h_1+h_2)}
\ee

As we can see, this method which is easier and trustable than
construction  of combinations of $osp(2|1)$ invariants\cite{q},
extendable to any corrolation functions.

We take $N=1$ superstring based on q-deformed Lie superalgebra $\uos$.
In this case, in order to construct the q-Ward identity, we should define the
co-product rule for the generators as follows:
\be
\Delta (L_0)=L_0\tp I+I\tp L_0,\hspace{0.5cm} \Delta(G_{\pm 1/2})=G_{\pm 1/2}\tp q^{L_0}
+q^{-L_0}\tp G_{\pm 1/2}.
\ee
It is easy to show that this co-product fulfills the axiom of co-associatively;
\be
(I\tp \Delta)\Delta=(\Delta\tp I)\Delta.
\ee
The difference operator realization is been given by (\ref{uos2}) with $H=2L_0$.
Then the q-Ward identity is given by
\be
\begin{array}{c}
\Delta G_{-1/2}\zz=\cr
[\d_{\t_1}+\f{\t_1}{z_1(q-q^{-1})}(M^{+1}_{1}-M^{-1}_{1})]q^h(1+(q^{1/2}-1)
\t_2\d_{\t_2})M^{+1}_2\zz\cr
+q^{-h}(1+(q^{-1/2}-1)
\t_1\d_{\t_1})M^{-1}_1
[\d_{\t_2}+\f{\t_2}{z_2(q-q^{-1})}(M^{+1}_{2}-M^{-1}_{2})]
\zz=0\cr
\Delta G_{+1/2}\zz=\cr
=[\f{q}{1+q}z_1 \dt{1} ( \q{2} \M{+1}{1} + q^{-2h-1}\M{-1}{1})
+\qq\t_1 (\q{2}\M{+2}{1}-\q{-2}\M{-2}{1})\cr
+\f{q}{(q+1)(q-q^{-1})}\t_1 (q^{2h-1}-\q{-2}
+ \q{-2}\M{-2}{1}-q^{2h-1}\M{2}{1})] \cr
\times \q{}( 1+ (q^{1/2}-1)\t_2 \dt{2})\M{+1}{2}\zz
+ \q{-} ( 1+ (q^{-1/2}-1)\t_1 \dt{1})\M{-1}{1}\cr
\times [\f{q}{1+q}z_2 \dt{2}(\q{2}\M{+1}{2}+ q^{-2h-1}\M{-1}{2})
+\qq \t_2 (\q{2}\M{+2}{2}-\q{-2}\M{-2}{2})\cr
+\f{q}{(q+1)(q-q^{-1})}\t_2 (q^{2h-1}-\q{-2}
+\q{-2}\M{-2}{2}-q^{2h-1}\M{+2}{2}]\zz=0
\end{array}
\ee
If we  solving the q-Ward  identity, we  have the following  expression for
the  two point
function
\be
\zz =
z_1^{-2h}\f{(\q{2}x;q^2)}{(q^{-2h}x;q^2)}
-\t_1\t_2
q^{-1/2}[2h]z_1^{-2h-1}\f{1}{1-q^{-2h-1}x}\f{(q^{2h+1}x;q^2)}{(q^{-2h+1}x;
q^2)}
\ee
where $ x=\f{z_2}{z_1}$ and
\be
(x;q^2)=\p_{n=0}^{\infty} (1-xq^{2n}).
\ee
In the limit $ q\rar 1$, the q-correlation function reduces
\be
\zz \rar (z_1-z _2+\t_1\t_2)^{-2h}
\ee
\noindent
{\bf Acknowledgment:}

The work of W.S.C. was supported in part by NON-DIRECTED RESEARCH FUND, Korea
Research Foundation, 1995, in part by the KOSEF through C.T.P. at Seoul
National
University and the present studies were supported in part by Basic Science
Research
Program, Ministry of Education of Korea, 1995(BSRI-94-2413).


\begin{thebibliography}{99}
\bibitem{aziz} A. Shafiekhani Mod. Phy. Lett. {\bf A9} (1994) 3273.

\bibitem{alx} A. Morozov and L. Vinet, {\it "Free Field Representation of Group
Element for Simple Quantum Group"} preprint ITEP-M3/94,CRM-2202 and
hep-th/9409093.

\bibitem{K} K. Kimura {\it "On Free Boson Representation of the Quantum
affine Algebra $U_q(\widehat{sl2})$"} Kyoto preprint, Dec.92.

\bibitem{LS} S. Lukyanov and S.L. Shatashvili, {\it"Free Field Representation
for the Classical Limit of Quantum Affine Algebra"} IASSNS-Hep-92/62 and
Ru-92/37, Sept. 92

\bibitem{SH} J. Shiraishi, {\it Phys. Lett.} {\bf A171} (1992) 243.

\bibitem{M} A. Matsuo, {\it Phys. Lett.} {\bf B308} (1993) 260.

\bibitem{A} A. Abada, {\it Mod. Phys. Lett.} {\bf A8} (1993) 715.

\bibitem{aos} H. Awata, S. Odake and J. Shiraishi, {\it "Free Boson
Representation of $U_q(\widehat{sl_3})$"} RIMS-920 and YITP/K-1017, May 93;
{\it "Free Boson Representation of $U_q(\widehat{sl_n})$"} to appear in C.M.P.

\bibitem{cft} G. Moore and N. Seiberg, Commun. Math. Phys. {\bf 123} (1989) 177;
A.B. Zamolodchikov, Rev. Math. Phys. {\bf 1} (1990) 197.
\bibitem{andre} A. Gerasimov, D. Lebedev, S. Khoroshkin, A. Mironov et al. {\it
"Generalized Hirota Equations and Representation Theory. 1. The Case of $SL(2)$
and $SL_q(2)$"}, preprint ITEP-M2/94, FINA/TD-5/94, NBI-HE-94-27,
{\it hep-th/9405011}.
\bibitem{frt}L.D. Faddeev, N.Yu. Reshtikhin and L.A. Takhtajan, Alg. Analiz.
{\bf 1} (1989) 1; R.B. Zhang, M.D. Gould and A.J. Bracken, J. Phys. {\bf A24}
(1991) 1185; C. Schwiebert {\it "Generalized Quantum Scattering"},
preprint RIMS and {\it hep-th/9412237}.

\bibitem{sm} H. Saleur, Nucl. Phys. {\bf B336} (1990) 363;
T. Deguchi and Y. Akutsu, J. Phys. {\bf A23} (1990) 143.

\bibitem{bgt} A.J. Backen, M.D. Gould and I. Tsohantjis, J. Math. Phys.
{\bf 34} (1993) 1654.

\bibitem{np} C. Nappi and E. Witten, Phys. Lett. {\bf 71} (1994) 3751
\bibitem{kt}A.C. Klimcik and A.A. Tseytlin, "{\it Duality invariant class of exact
string background}", hep-th/{\it 9311012}
\bibitem{ors} D.I. Olive, E. Rabinovic and A. Schwimmer, "{\it A class of string
background as a semiclassical limit of WZW models}", hep-th/{\it 9311081}
\bibitem{s} K. Sfetsos, "{\it Gauging a non-semi-simple WZW model}",
hep-th/{\it 9311010}
\bibitem{m} N. Mohammadi, Phys. Let. {\bf B325} (1994) 371.
\bibitem{fs} J.M. Figueroa-O'Farril and S. Stanciu, Phys. Let. {\bf b327} (1994)
 40.
\bibitem{q} Z. Qiu, Nucl. Phys. {\bf B270} [FS16] (1986) 205
\bibitem{ks} E.B. Kiritsis and G. Siopsis, Phys. Lett. {\bf B184} (1987) 353
\bibitem{ch}P.P. Kulish, Kyoto preprint RIMS-615 (1988);
P.P. Kulish and N.Yu. Reshetikhin, Lett. Math. Phys. {\bf 18}
(1989) 143; M. Chaichian and P. Kulish, Phys. Lett. {\bf B 235} (1990) 72.
\bibitem{bpz}A.M. Belavin, A.M. Polyakov and A.B. Zamolodchikov, Nucl. Phys.
 {\bf B241} (1984) 333.
\end{thebibliography}
\end{document}